\documentclass[a4paper]{article}
\usepackage{ASVspoof5}
\usepackage{epsfig,amssymb,amsmath}
\usepackage{hyperref}
\usepackage{xcolor}
\usepackage{tabularx,lipsum}
\usepackage{ctable}
\usepackage{float}
\usepackage{balance,subcaption}
\usepackage{comment}
\usepackage{svg}
\usepackage{censor}
\usepackage{url,cite}
\hypersetup{
    colorlinks=true,
    linkcolor=purple,
    filecolor=magenta,      
    urlcolor=purple,
}
\ninept

\title{Malacopula: adversarial automatic speaker verification attacks\\ using a neural-based generalised Hammerstein model}

\name{John Doe}

\makeatletter
\def\name#1{\gdef\@name{#1\\}}
\makeatother
\name{{\em Massimiliano Todisco$^1$, Michele Panariello$^1$, Xin Wang$^2$,}\\
     {\em Héctor Delgado$^3$, Kong Aik Lee$^4$, Nicholas Evans$^1$}}

\address{$^1$EURECOM, Sophia Antipolis, France, $^2$National Institute of Informatics, Japan\\$^3$Microsoft, Spain, $^4$The Hong Kong Polytechnic University, Hong Kong\\
{\small \tt $^1$\textit{firstname}.\textit{lastname}@eurecom.fr} }
\begin{document}
\maketitle

\begin{abstract}
We present Malacopula, a neural-based generalised Hammerstein model designed to introduce adversarial perturbations to spoofed speech utterances so that they better deceive automatic speaker verification (ASV) systems. Using non-linear processes to modify speech utterances, Malacopula enhances the effectiveness of spoofing attacks. The model comprises parallel branches of polynomial functions followed by linear time-invariant filters. The adversarial optimisation procedure acts to minimise the cosine distance between speaker embeddings extracted from spoofed and bona fide utterances. Experiments, performed using three recent ASV systems and the ASVspoof 2019 dataset, show that Malacopula increases vulnerabilities by a substantial margin. However, speech quality is reduced and attacks can be detected effectively under controlled conditions. The findings emphasise the need to identify new vulnerabilities and design defences to protect ASV systems from adversarial attacks in the wild.
\end{abstract}

\section{Introduction}
The performance of automatic speaker verification (ASV) systems has improved remarkably in recent years. 
The pioneering x-vector approach~\cite{snyder2018xvector} laid the foundation for more recent and robust systems including ECAPA~\cite{ECAPA}, CAM++~\cite{CAM++}, and ERes2Net~\cite{ERes2Net} which consistently outperform their predecessors in various benchmarks~\cite{JAKUBEC2024107232}.

Despite these technological advances, ASV systems remain vulnerable to spoofing attacks implemented using, e.g., text-to-speech synthesis and voice conversion techniques. These attacks have become increasingly sophisticated, capable of producing spoofed speech which is generally indistinguishable from bona fide speech and which effectively compromise ASV reliability. Nonetheless, there is evidence that recent ASV systems have some natural defensive capabilities against spoofing attacks~\cite{10415203}. Natural defences can also be supplemented using auxiliary spoofing and deepfake detection solutions~\cite{10155166}.

While the study of spoofing and the development of detection solutions has attracted broad attention, a new threat has emerged in the form of adversarial attacks, e.g.~\cite{audio_adv_attacks_carlini}. 
These are implemented using adversarial training which, in the context of ASV, can be used by a fraudster to introduce noise—sometimes imperceptible or easily mistakable for real environmental sounds—to deceive the classifier and provoke a higher rate of false alarms/acceptances. 

In recent work~\cite{malafide}, we showed how adversarial training techniques can be used to design a simple linear time-invariant (LTI) filter, named Malafide, which compromises the reliability of even state-of-the-art spoofing and deepfake detection solutions.  In this paper we report our work to evaluate the robustness of ASV systems to the same form of adversarial attacks.  We introduce Malacopula,\footnote{\textit{Mala copula} is Latin for "bad connection" or "bad union." It signifies an undesirable or improper association between elements.} a neural-based generalised Hammerstein model~\cite{00207170500096666} designed specifically to compromise ASV system reliability through the introduction of adversarial perturbations to a test speech utterance.  
Unlike Malafide, Malacopula supports the modification of not only amplitude and phase but also frequency components in non-linear fashion, a crucial benefit for voice cloning.

Malacpolua acts as a post-processing filter to increase ASV system vulnerabilities to spoofing attacks.  Tuned to the spoofing attack and speaker identity, the Malacopula filter is optimised independently of the utterance and input duration, requiring the optimisation of only a small number of filter coefficients, in similar fashion to Malafide~\cite{malafide}.

\section{Literature Review}
Adversarial attacks were originally introduced for image processing tasks~\cite{adv_attacks_goodfellow}, but have since been applied to the speech domain, particularly focusing on automatic speech recognition (ASR)\cite{audio_adv_attacks_carlini,neekhara19b_universal} and spoofing/automatic speaker verification (ASV) systems\cite{xie2020real,universal_perturbations_CM,adv_kassis}.

Early strategies involved generating adversarial noise specific to each utterance, drawing inspiration from image processing techniques~\cite{universal_perturbations_image}. These strategies adapted universal adversarial perturbations to various audio tasks, including automatic speech and speaker recognition~\cite{li2024initial,universal_perturbations_ASR,unviersal_perturbations_ASV}. A common theme among these methods is the iterative optimisation of adversarial perturbations across multiple data samples.

Initial research~\cite{adv_attacks_goodfellow, adv_kassis} primarily explored adversarial examples as additive noise and their ability to transfer to unseen scenarios.
The study in~\cite{universal_perturbations_CM} explored universal perturbations against spoofing and deepfake countermeasure (CM) systems.
This method targets both CM and ASV subsystems independently of specific attacks.
However, it requires the generation of a different array of adversarial noise for each utterance, which results in a high computational effort.
Moreover, the variable length of speech utterances is a constraint upon the generation of adversarial noise, rendering these attacks impractical in real-world scenarios.

Malafide~\cite{malafide} introduced an adversarial technique utilising linear time-invariant (LTI) filters applied in real-time to spoofed utterances through time-domain convolution. Unlike traditional methods, Malafide filters are optimised independently of the input utterance and its duration, tailored specifically to the underlying spoofing attack. This method requires the optimisation of only a small number of filter coefficients, thereby offering greater flexibility in its application.

Our approach takes a different path by enhancing specific spoofing attacks and targeting specific speakers to increase the threat to ASV systems. We operate under the assumption that the spoofing attack effectively manipulates the ASV subsystem.

In contrast to prior research, our method involves the use of adversarial, non-linear filters using a generalised Hammerstein model, commonly used for the identification of non-linear systems. Malacopula can be applied in real-time to a spoofed utterance via time-domain convolution operations, specifically targeting a particular speaker and the underlying attack algorithm.

\section{Generalised Hammerstein Model}

The generalised Hammerstein model is a prominent framework in signal processing, employed to identify non-linear dynamic systems. The model combines a static non-linear component with a linear dynamic component, enabling detailed representation and manipulation of complex signal characteristics.

The structure of the Generalised Hammerstein Model typically comprises two main elements: a non-linear transformation followed by a linear time-invariant (LTI) filter. 
Mathematically, the model can be expressed as:
\begin{equation}
    y[n] = \sum_{k=1}^{K} \sum_{i=0}^{L-1} h_k[i] \phi_k(x[n-i])
\end{equation}

\noindent where \( y[n] \) is the output signal, \( x[n] \) is the input signal, \( \phi_k(x[n-i]) \) represents the static non-linear transformation, \( h_k[n] \) represents the impulse response of the LTI filter for the \( k \)-th branch, \( L \) denotes the memory length, \( K \) is the number of parallel branches and \( n \)  represents the discrete sample index.
The non-linear transformation captures the non-linearities of the input signal, often modelled using polynomials as functions of the input signal: \(\phi_k(\cdot) = (\cdot)^k\). 
The versatility and computational efficiency of the generalised Hammerstein model have facilitated its successful application to the modelling of non-linear systems across various fields, including audio processing, acoustics, and mechanical vibrations~\cite{grimm2016hybrid, 1599765, 4124576, 9424600}. 

Polynomial Hammerstein models have been employed to characterise and model non-linear loudspeakers using empirically measured Volterra kernels~\cite{6953353}. Results show the potential of the approach in estimating reliable non-linear models which accurately predict the response to complex real-speech inputs.
For the collection of the ASVspoof 2019 physical access (PA) databases~\cite{ASVspoof2019}, the generalised Hammerstein model was utilised to model and simulate loudspeaker artefacts which often stem from non-linear behaviour. Both linear and non-linear characteristics 
accurately simulating the distortions introduced by loudspeakers.
RawBoost~\cite{RawBoost} leverages the generalised Hammerstein model within a machine learning framework primarily for augmentation purposes rather than system identification. RawBoost simulates a wide range of signal distortions, thereby improving the robustness and generalisation of machine learning models trained with augmented data. RawBoost was developed specifically for the detection of spoofing and deepfakes  in the wild but has also been used effectively in other speech-related applications~\cite{10605761}.

\begin{figure*}[!t]
    \centering
    \includegraphics[width=\textwidth]{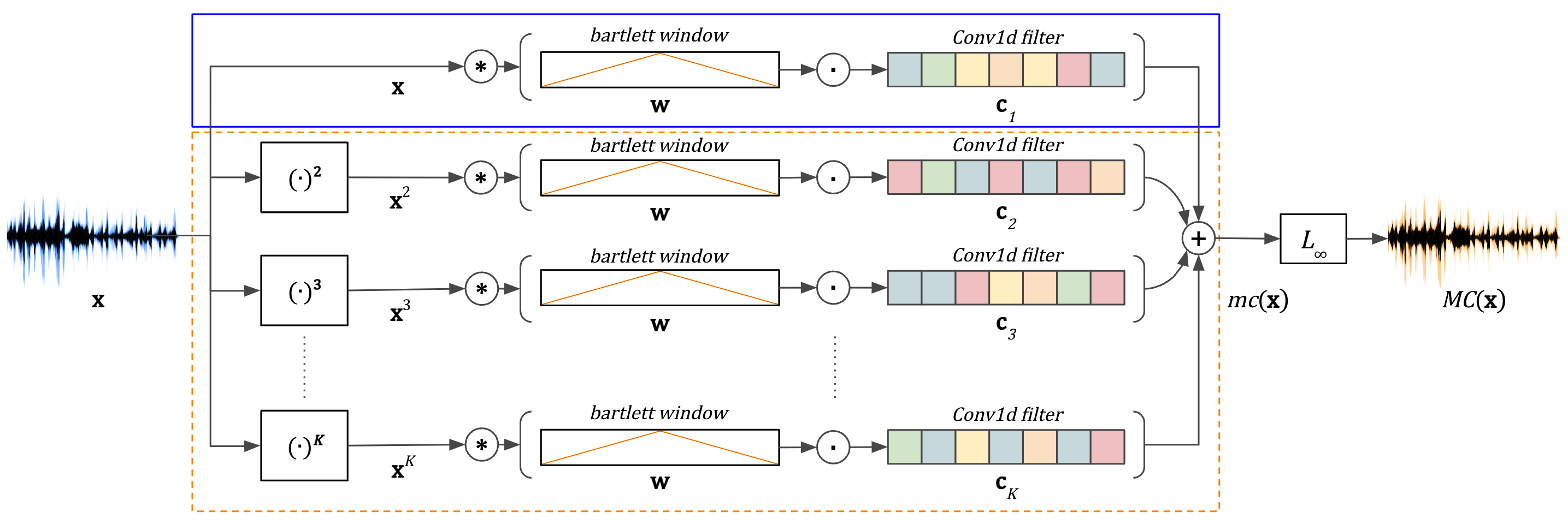}
    \caption{Malacopula filter architecture based on the generalised Hammerstein model. The blue box represents the linear component, while the the orange dashed box represents the non-linear filter components.}
    \label{fig:Malacopula}
\end{figure*}

\begin{figure*}[!t]
    \centering
    \includegraphics[width=\textwidth]{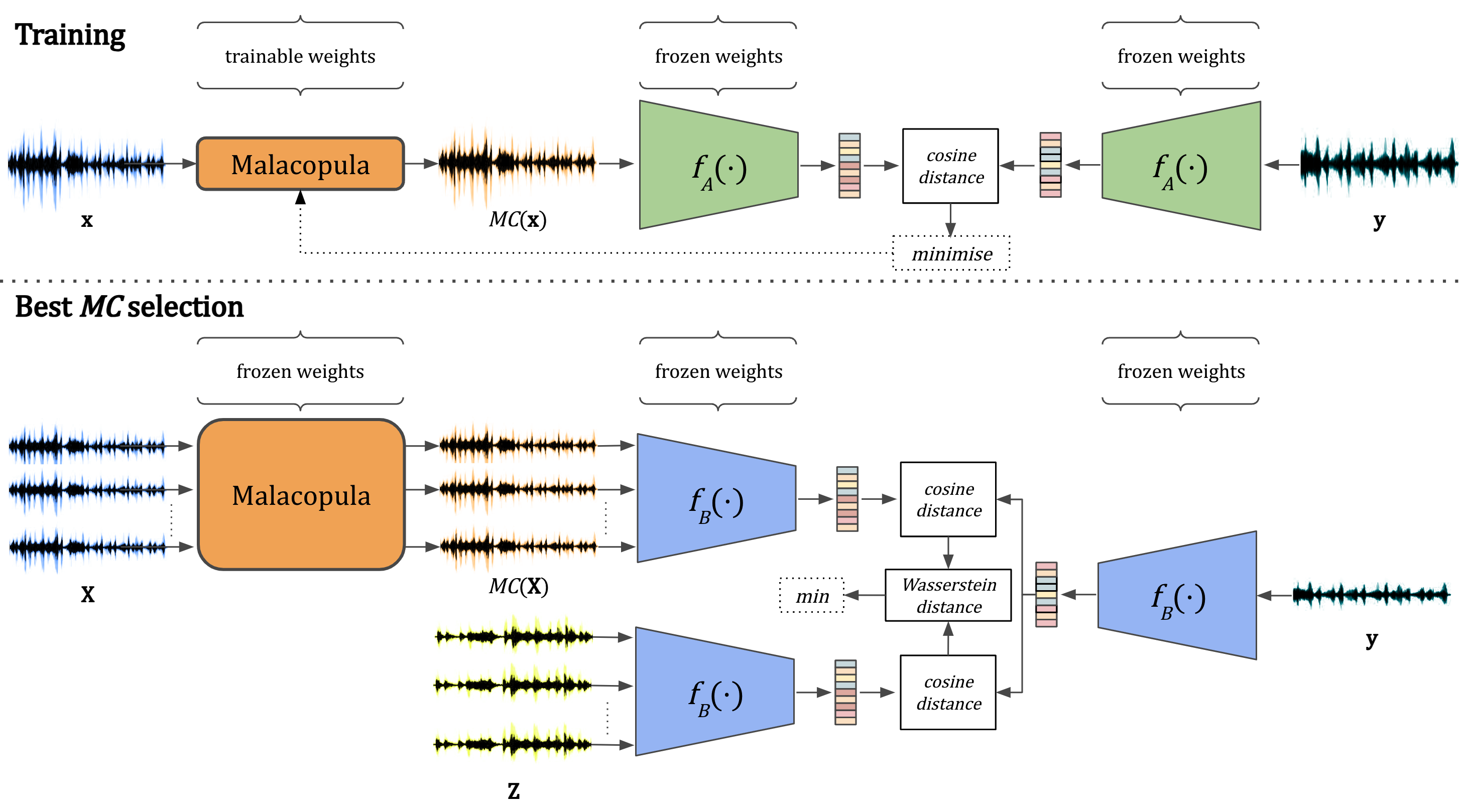}
    \caption{ 
    During training, Malacopula filters are optimised with the speaker embedding extractor \(f_A(\cdot)\) as denoted by Equation~\ref{eq:objective}. To ensure generalisation across different speakers, the best Malacopula filter is selected using another speaker embedding extractor \(f_B(\cdot)\). The selection is based on the  minimum Wasserstein distance between the following two score distributions: (i)~the cosine distance between spoofed utterances processed by the Malacopula filter \(MC(\mathbf{X})\) and bona fide enrolment utterances \(\mathbf{y}\), and (ii)~the cosine distance between bona fide target utterances \(\mathbf{Z}\) and bona fide enrolment utterances \(\mathbf{y}\). If multiple enrolment utterances are available, we use the average enrolment embedding.
}
    \label{fig:malatrain}
\end{figure*}

\section{Malacopula}
The generalised Hammerstein model offers a powerful method to manipulate  multiple characteristics of a speech signal, including the modification of amplitude and  phase, but also frequency components in non-linear fashion.
This capability can be exploited by malicious actors to create adversarial perturbations in order to deceive ASV systems.
In the following we describe the implementation of the Malacopula filter.

\subsection{Malacopula filter architecture}
The Malacopula filter structure is shown in Fig.~\ref{fig:Malacopula}. It is composed of \(K\) parallel branches, which represent the non-linear depth, each involving a linear filter \(\mathbf{c}\) of length \(L\) modulated by a Bartlett window \(\mathbf{w}\).\footnote{A Bartlett window, also known as a triangular window, is used in signal processing to reduce the side lobes of the filter response, which helps in minimising the spectral leakage. This window is chosen because it provides a trade-off between the width of the main lobe and the level of side lobes, making it suitable for applications where both frequency resolution and dynamic range are important.} Each branch processes the input signal \(\mathbf{x}\) by a \(k\)-th non-linear, static power polynomial function. The filter operates entirely in the discrete time domain using convolution operations.

Mathematically, the filter is defined by:
\begin{equation}
    mc_{K,L}(\mathbf{x})= \sum_{k=1}^{K}\left[ \mathbf{x}^k \ast \left( \mathbf{w} \odot \mathbf{c}_{k,L} \right) \right]
\end{equation}

\noindent where \(\ast\) denotes the convolution operator, and \(\odot\) represents the Hadamard product.

Additionally, a normalisation layer using the $L_\infty$
norm is applied after the summation operator to produce the output:

\begin{equation}
MC(\mathbf{x}) = \frac{mc(\mathbf{x})}{|mc(\mathbf{x})|_{\infty}}
\end{equation}

\begin{figure*}[!t]
    \centering
    \includegraphics[width=0.96\textwidth]{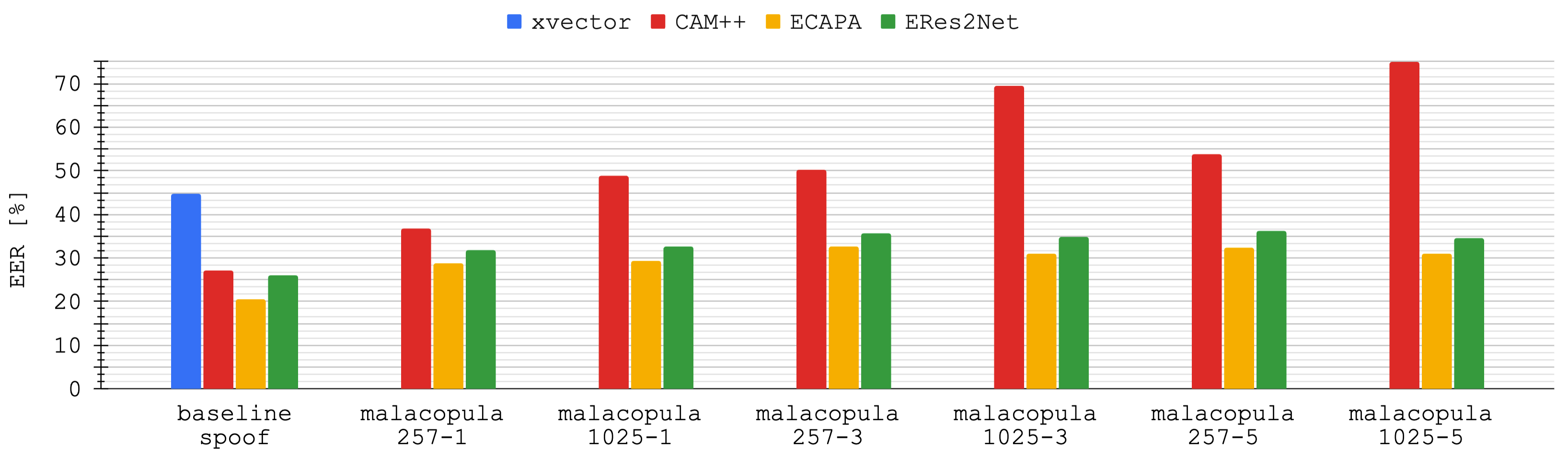}
    \caption{Pooled spf-EER for baseline spoof and Malacopula filtered spoof attacks for four different ASV systems.}
    \label{fig:pooled_Malacopula_ASV_results}
\end{figure*}

\subsection{Adversarial Optimisation Procedure}
The Malacopula optimisation procedure is illustrated in Fig.~\ref{fig:malatrain}.
Each filter is trained independently for a given speaker \(s\), and a spoof utterance \(\mathbf{x}\) generated with spoofing algorithm \(a\), and a bona fide enrolment utterance \(\mathbf{y}\).
The neural-based generalised Hammerstein model minimises the following objective function:
\begin{equation}
    \min_{\mathbf{c}^{(s,a)}_{K,L}} \left[ 1 - CS\left( f_A\left(MC^{(s,a)}_{K,L}(\mathbf{x})\right), f_A\left(\mathbf{y}\right) \right) \right]
    \label{eq:objective}
\end{equation}

\noindent where 
\(f_A(\cdot)\) denotes the pre-trained speaker embedding extractor, and \(CS(\mathbf{A}, \mathbf{B}) = \frac{\mathbf{A} \cdot \mathbf{B}}{\|\mathbf{A}\| \|\mathbf{B}\|}\) is the cosine similarity between embeddings \(\mathbf{A}\) and \(\mathbf{B}\).
The objective function aims to minimise the cosine distance between the speaker embeddings of the modified input signal \(MC(\mathbf{x})\) and the target signal \(\mathbf{y}\), ensuring that the adversarial perturbations are effective.

To ensure that the adversarial attacks generalise well across different speaker embeddings, the best Malacopula filter is selected using another speaker embedding extractor \(f_B(\cdot)\).
Filter selection is based on the minimum Wasserstein distance\footnote{The Wasserstein distance is chosen because it provides a robust measure of the similarity between two probability distributions by considering the 'cost' of transforming one distribution into another. This property is particularly useful in evaluating the similarity between bona fide and spoof scores, as it captures differences in both the shape and location of the distributions, ensuring that adversarial examples remain close to bona fide data. Unlike the Equal Error Rate (EER), which only considers the point where the false acceptance rate equals the false rejection rate, the Wasserstein distance evaluates the entire distribution, offering a more comprehensive assessment of distribution similarity. The EER may not effectively capture the nuances of distributional changes caused by adversarial perturbations.} computed across all training iterations, and between the following two score distributions: (i) the cosine distance between embeddings extracted from spoofed utterances processed by the Malacopula \(MC(\mathbf{X})\) and those extracted from  bona fide enrolment utterances \(\mathbf{y}\), and (ii) the cosine distance between embeddings extracted from bona fide target utterances \(\mathbf{Z}\) and bona fide enrolment utterances \(\mathbf{y}\). If multiple enrolment utterances are available, we use the average of their embeddings as the final enrolment embedding. Here \(\mathbf{X}\) and \(\mathbf{Z}\) are batches of speech utterances. This approach hence ensures that adversarial examples are sufficiently similar to the voice of the original speaker. Specifically, we employ a signed Wasserstein distance to incorporate not only the magnitude but also the direction of the distance. A positive Wasserstein distance is considered if the median of the distribution of spoof scores exceeds that of the target bona fide scores.

\begin{figure*}[!t]
    \centering
    \includegraphics[width=0.96\textwidth]{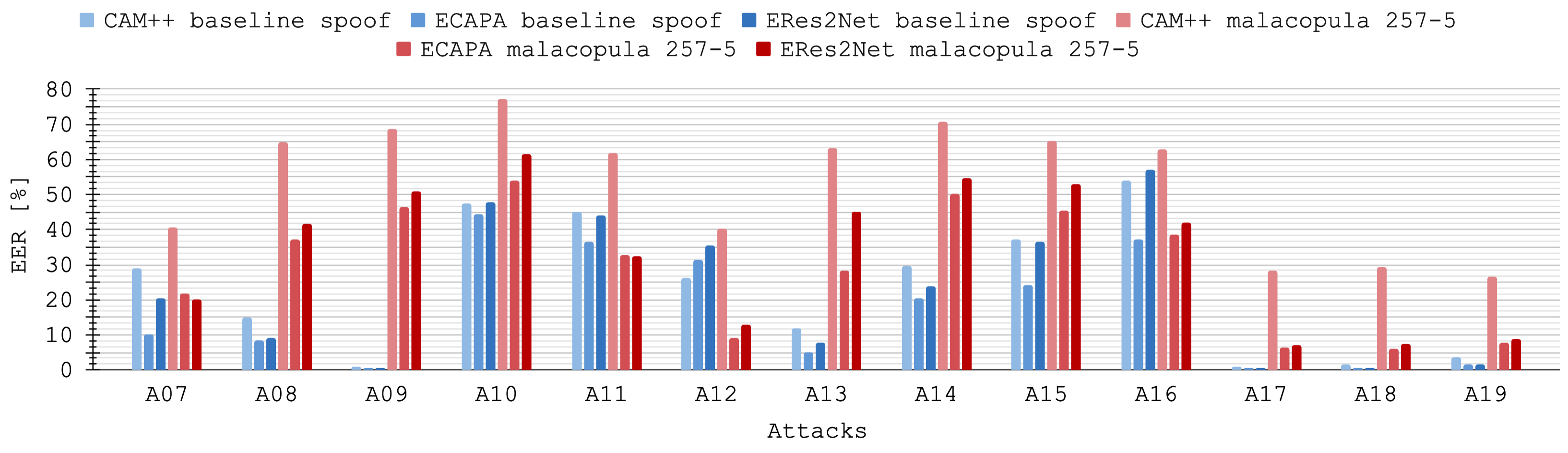}
    \caption{spf-EER per attacks for baseline spoof and Malacopula 257-5 filtered spoof attacks of three ASV systems.}
    \label{fig:attack-wise_Malacopula_ASV_results}
\end{figure*}

\begin{figure*}[!t]
    \centering
    \includegraphics[width=0.96\textwidth]{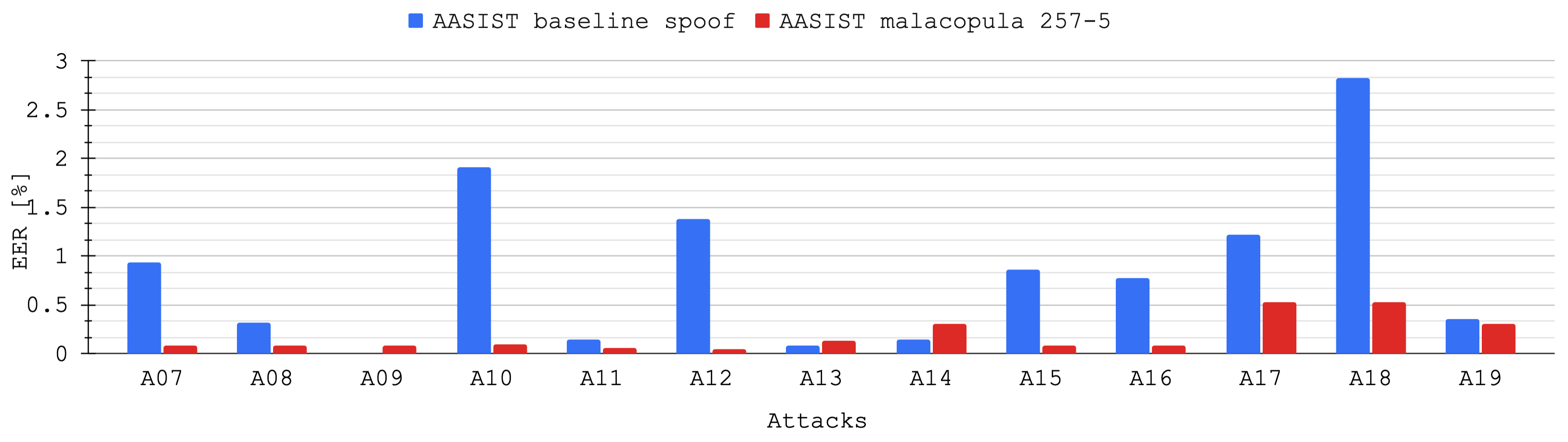}
    \caption{spf-EER per attacks for baseline spoof and Malacopula 257-5 filtered spoof attacks of three ASV systems.}
    \label{fig:AASIST_results}
\end{figure*}

\begin{figure}[!t]
    \centering
    \includegraphics[width=\columnwidth]{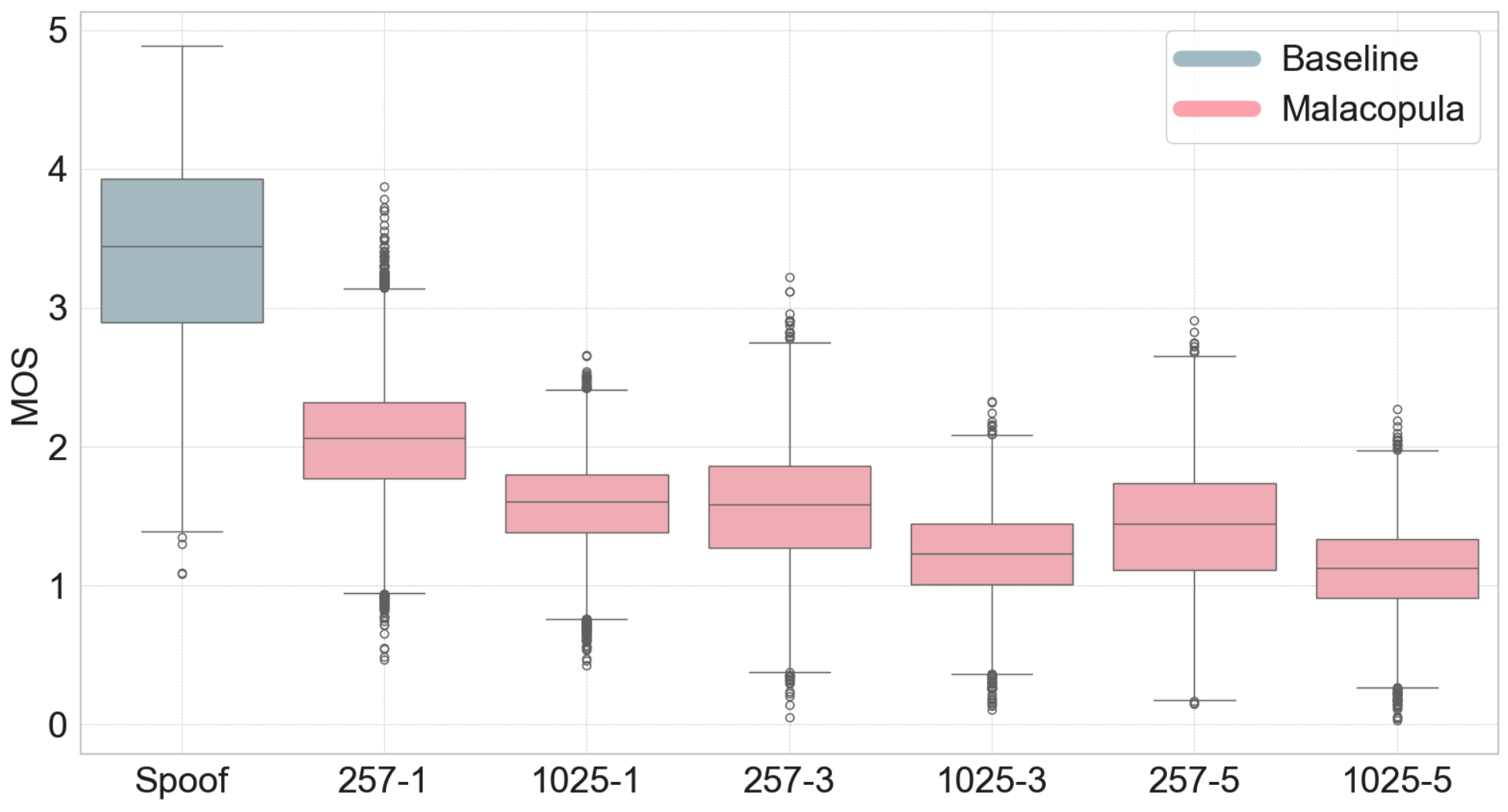}
    \caption{MOS distributions for baseline spoof and Malacopula filtered spoof.}
    \label{fig:mos_results}
\end{figure}

\section{Experimental Setup}
We use three distinct ASV systems, each with unique structural and functional characteristics. They are: CAM++ for training; ECAPA for validation; ERes2Net for testing. 
By employing three different ASV systems, we are able to test the transferability of Malacopula attacks across different ASV architectures and embedding extraction methodologies.
Below, we provide a brief descriptions of each system.

The \textbf{CAM++}~\cite{CAM++} system consists of a front-end convolution module and a densely connected time delay neural network (D-TDNN) backbone and the extraction of a 512-dimensional speaker embedding. Each D-TDNN layer includes a context-aware masking (CAM) module. CAM++ employs multi-granularity pooling to capture discriminative speaker characteristics, enhancing its ability to differentiate between speakers effectively.

The \textbf{ECAPA}~\cite{ECAPA} system uses the TDNN architecture, which incorporates 3 SE-Res2Block modules to extract a 192-dimensional speaker embedding. This structure leverages squeeze-and-excitation (SE) mechanisms to enhance feature representation and improve performance in speaker verification tasks.

The \textbf{ERes2Net}~\cite{ERes2Net} system addresses the limitations of the traditional Res2Net architecture by integrating local and global feature fusion  to extract a 192-dimensional speaker embedding. This integration allows ERes2Net to capture both detailed and holistic patterns in the input signal, enhancing its capability to recognise speaker-specific characteristics.

\subsection{Database, protocols and filter optimisation}
All experiments were conducted using the ASVspoof 2019 logical access (LA) dataset~\cite{ASVspoof2019}. It contains spoofing attacks generated with a set of algorithms labelled A01 to A19.
Attacks A01 to A06 are contained in both the \emph{training} and \emph{development} partitions, while A07 to A19 are contained only in the \emph{evaluation} partition.
Training and development partitions relate to the realm of a defender whose role is to train and develop spoofing and deepfake detectors. 
In contrast, the test partition contains data in the realm of the attacker.
Speaker and attack-specific filters are hence trained according to~\eqref{eq:objective} using the test partition, i.e.\ using A07 to A19 spoofing attack data. 
We stress that, in contrast to usual practice, the use of \emph{test} data for \emph{training} purposes is acceptable in this case; the attacker is not bound by experimental protocols and can use test data in any reasonable way which is to their advantage.

Malacopula filters are trained using spoofed and bona fide utterances sourced from the \textit{test} data partition and using CAM++ 
for $f_A(\cdot)$ 
and ECAPA 
as $f_B(\cdot)$, 
while testing is performed using ERes2Net.
The setup reflects a scenario in which filters are trained by an attacker offline and then used to implement online/real-time attacks, e.g.\ in a logical access or telephony scenario.

\subsection{Implementation}
The objective function \eqref{eq:objective} is optimised with Adam~\cite{kingma2014adam}.
Filters are optimised for $60$ epochs with a batch size of $12$.
ASV model weights are kept frozen during optimisation.
We used two filter lengths \(L \in \{257,1025\}\) and three filter depths \(K \in \{1,3,5\}\) to explore the balance between optimisation of~\eqref{eq:objective}, attack success and the preservation of speech quality.
Our specific implementation, along with audio examples, is available as open-source and can be used to reproduce our results under the same GPU environment.\footnote{\url{https://github.com/eurecom-asp/malacopula}}

\subsection{Metrics}
All results are obtained using the standard SASV evaluation protocol~\cite{jungsasv2022} and are expressed in terms of spf-EERs computed using target (positive class) and spoofed (negative class) utterances.

\section{Experimental Results}
Results presented in Fig.~\ref{fig:pooled_Malacopula_ASV_results} show pooled spf-EERs for the three ASV systems, comparing baseline spoof results with those using Malacopula filters of different length and depth. For comparison with a legacy ASV system, we include performance for  an x-vector system among the baseline spoof results to
show that modern systems are less vulnerable. For Malacopula attacks, the vulnerability increases universally, and more significantly for CAM++ which is used for training.
Still, spf-EERs are higher for ECAPA and ERes2Net systems than for the baseline condition. 
This shows that Malacopula filters exhibit some generalisation across different ASV architectures. 

Fig.~\ref{fig:attack-wise_Malacopula_ASV_results} show a performance comparison 
for baseline spoof and Malacopula 257-5 filters for the three systems — CAM++, ECAPA, and ERes2Net — for all thirteen underlying spoof attacks. For certain attacks, such as A09, Malacopula leads to a significant increase in the vulnerability of all three systems. 
However, Malacopula exhibits lower performance against certain already-effective attacks, such as A12. For attacks A17, A18, and A19, which are all voice conversion based attacks, a similar increase in vulnerability is observed.

Overall, results show that Malacopula filters provoke increased vulnerabilities across the three ASV systems and attack scenarios. This underscores the importance of continuous improvement and adaptation in ASV system defences to maintain robustness against evolving adversarial techniques.

Fig.~\ref{fig:AASIST_results} shows the impact of Malacopula 257-5 filters upon the popular ASSIST~\cite{jung2022aasist} spoof and deepfake detection system. Results are shown in terms of spf-EERs for baseline spoof (blue bars) and Malacopula attacks (red bars). 
When the utterances are processed by Malacopula, AASIST performance \emph{improves} almost universally.  Only for A13 and A14 are spf-EERs higher, albeit only very marginally, and are in any case still low. These results indicate that Malacopula  attacks are easily detectable, reinforcing the need for dedicated detection solutions in order to protect ASV systems from manipulation. 

Fig.~\ref{fig:mos_results} illustrates the impact on speech quality measured in terms of the mean opinion score (MOS) for various Malacopula configurations. All scores were estimated automatically using the method described in~\cite{mos_junichi}. 
MOS distributions are shown for the baseline spoof and Malacopula attacks \((L,K)\). As expected, distributions for the baseline spoof attacks are generally higher, with distribution modes of around 3 and 4, For Malacopula attacks, distribution modes are between 1 and 2. Variations in speech quality caused by Malacopula attacks are attributed to the use of more or less aggressive filters, where smaller values of \(L\) and \(K\) cause less degradation.

However, the controlled conditions under which ASVspoof 2019 source data was initially collected, do not reflect factors such as background or channel noise, which typify conditions in the wild and which may influence results. 
Perturbations introduced by Malacopula themselves resemble background or channel noise. This suggests the need for further investigations to verify detection performance in more realistic acoustic conditions and scenarios. 

\section{Conclusions}
In this paper, we introduce Malacopula, an adversarial perturbation model in the form of generalised Hammerstein framework, which acts upon a speech utterance in order to exaggerate and exploit the vulnerabilities of automatic speaker verification systems to spoofing and deepfake attacks. 
Malacopula extends the capabilities of previous models, enabling more effective manipulation of the amplitude, phase, and frequency components of speech signals in non-linear fashion.

Experiments, performed using the ASVspoof 2019 dataset show that Malacopula significantly increases the vulnerability of CAM++, ECAPA, and ERes2Net ASV systems to spoofing and deepfake attacks. The cross-system training and evaluation nature of the experiments underscores the robustness and transferability of Malacopula attacks, highlighting the potential threat in real-world scenarios.

Despite the power of Malacopula in increasing the threat of spoofing attacks, our analysis reveals that the resulting perturbations reduce speech quality, as reflected by lower mean opinion scores. Reassuringly, though, spoofing and deepfake detection systems like AASIST are capable of detecting Malacopula attacks.
However, we acknowledge that our current work shows only that attacks are detected effectively under controlled conditions. 
This suggests the need for further investigations to determine whether the same defences remain robust in unconstrained scenarios.
Our findings highlight the importance of continuing the hunt for new vulnerabilities and efforts to tackle them so as to ensure the reliability of ASV systems in the wild. 

\section{Acknowledgements}
This work is supported with funding received from the French Agence Nationale de la Recherche (ANR) via the BRUEL (ANR-22-CE39-0009) and COMPROMIS (ANR-22-PECY-0011) projects.

\bibliographystyle{IEEEbib}
\balance
\bibliography{ASVspoof5_BibEntries}

\end{document}